\documentclass[12pt,english]{article}
\usepackage[T1]{fontenc}
\usepackage[latin9]{inputenc}
\usepackage[letterpaper]{geometry}
\usepackage{setspace}
\usepackage{amssymb}
\usepackage{esint}
\usepackage{babel}

\begin{document}

\title{Curvature independence of statistical entropy}
\author{  Judy Kupferman \\   Ben-Gurion
University, Beer-Sheva, 84105 Israel \\ }
\date{}
%\maketitle

\author{ Judy Kupferman\\  Ben-Gurion
University, Beer-Sheva, 84105 Israel }
\maketitle
\begin{abstract}
We examine the statistical number of states, from which statistical
entropy can be derived, and we show that it is an explicit function
of the metric and thus observer dependent. We find a constraint on
a transformation of the metric that preserves the number of states
but does not preserve curvature.  In showing exactly how curvature independence arises in the conventional definition of statistical entropy, we gain a precise understanding of the direction in which it needs to be redefined in the treatment of black hole entropy.
\end{abstract}

\section{Introduction}

Black hole entropy is discussed in a number of contexts: thermodynamics
and statistical mechanics \cite{BCH,Bekenstein,Gibbons hawking,'t Hooft,Wald review},
quantum entanglement \cite{Bombelli,Srednicki,Plenio}, spacetime
symmetries \cite{Wald 1,wald iyer,Carlip,Silva} and more. It is not
clear whether entropy in all these contexts refers to the same entity,
although they are frequently taken to be equivalent. The fact that
black holes obey thermodynamic type laws is still not understood;
for example we do not know what degrees of freedom the entropy represents.
If the different versions of entropy do not coincide, the confusion
increases. In order to shed some light on the matter, we examine statistical
entropy in curved space, and we ask whether statistical entropy can
be related to the curvature of spacetime. Our main motivation is an attempt to introduce a little more clarity into the discussion of entropy in the black hole context, by providing an unequivocal description of the curvature independence of statistical entropy as it is conventionally defined in the literature.

The Unruh effect shows that an accelerated observer sees a thermal
bath of particles, while an observer in Minkowski space sees vacuum
\cite{Unruh} so clearly the statistical entropy in the two
cases will be different.  Both Minkowski and Rindler metric have the
same (vanishing) curvature, so it appears that the entropy must be
observer dependent and not a function of curvature.  On the other hand
Wald's Noether charge entropy \cite{Wald 1,wald iyer} is defined
in terms of the curvature tensor, \begin{equation}
S_{Wald}=2\pi\intop_{\Sigma}\frac{\delta L}{\delta R_{abcd}}\epsilon_{ab}\epsilon_{cd}\end{equation}
where the functional derivative is taken viewing the Riemann tensor
as a field independent of $g_{ab}$, and $\epsilon_{ab}$ is the binormal
to the bifurcation surface. Thus it appears to be different from the
entropy defined in statistical mechanics. The question is whether
for any curvature (not just in the absence of curvature) the number
of states is observer dependent, or whether there is a possible dependence
on curvature.  The textbook definition of statistical entroy originally assumed flat space, where the choice of vacuum is unambiguous. In curved space this is not the case. Thus it would appear that this definition is not suitable for treatment of black hole entropy. We will show technically where and how curvature independence arises in the conventional definition. This will give a precise understanding of the direction in which it needs to be redefined.

In this paper we examine the statistical number of states of matter
in a general curved space metric. Our motivation is understanding
of black hole entropy; however, specialisation to the black hole metric
includes further issues to be dealt with in future work. The relationship
of the entropy of matter outside the horizon to the entropy of the
black hole is not clear. 't Hooft \cite{'t Hooft} calculated the
statistical entropy of a scalar field in the black hole metric, where
his motivation was to reconcile black hole physics with quantum mechanics.
At the time of writing, black holes were understood to be in a quantum
mechanically mixed state, and 't Hooft attempted to describe them
as pure states resembling ordinary particles. Thus black holes inhabit
an extension of Hilbert space with an according Hamiltonian. This
system is sensitive to observer dependence: the free falling observer
perceives matter, and 't Hooft writes that it is this matter which
he considers in this paper. The distinction between presence and absence of matter
is assumed to be observer dependent when considering coordinate transformations
with a horizon. Another view on the relationship between the statistical
entropy of a field outside the horizon and the black hole is the idea
that the horizon entropy arises from the microscopic structure of
spacetime and that the matter fields inherit the entropy as material
kept in a hot oven inherits the temperature \cite{paddy kolekar ref 13,Kolekar}.
Yet another view takes into account the fact that the entanglement
entropy of a bipartite system, which expresses the quantum correlations
between its subsystems, is equal to the statistical entropy of a subsystem
if it is in a thermal state \cite{Kabat,kabat 3 israel,JudyRamy},
thus statistical entropy of fields outside the horizon may equal entanglement
entropy of the black hole system%
\footnote{The statistical entropy, as discussed in this paper, is fundamentally
different than the usual notion of quantum entanglement entropy. This
distinction is further clarified in the concluding section.%
}. In this paper we do not discuss these issues. We focus only on the
curvature independence of statistical entropy in a general curved
space, as a first small step in clarifying the relation of the different
concepts of black hole entropy.

We will show that the number of states from which statistical entropy
is derived is an explicit function of the metric. Since the curvature
derives from the metric, it would seem that the number of states is
related to curvature. However we find that for certain transformations
of the metric, the number of states is preserved. These transformations
do not preserve curvature. Therefore the number of states does not
depend on curvature. This is shown only for a diagonal metric, but
it serves as a counter example showing that in the most general case
the number of states is not a function of the spacetime curvature
scalar.

This paper is organized as follows. First we establish the definition
of the number of states, and methods of calculating the volume of
phase space. We then ask under what conditions transformation of the
metric will leave this volume invariant. We obtain a general transformation
of any diagonal metric which displays a clear constraint on the preservation
of the number of states. We examine characteristics of this transformation
and look for a possible relationship to curvature. We find that in
general it need not preserve curvature. That is, the number of states
and thus the entropy will remain the same for systems with different
curvature. This is shown for a diagonal metric in a static spacetime,
but serves as a counter example showing that in general entropy is
not dependent on curvature.

\section{Definition of number of states}

In classical thermodynamics the number of states of a nonrelativistic
system is defined as follows: Take an integral over the volume of
phase space ($d^{3}xd^{3}p$), restrict it to values of momenta which
fit the energy eigenvalues of the system and the number of states
is\begin{eqnarray}
N & = & g\int d^{3}x\int\frac{d^{3}p}{\left(2\pi\hbar\right)^{3}}\nonumber \\
 & = & gV\int\frac{d^{3}p}{\left(2\pi\hbar\right)^{3}}\end{eqnarray}
where $g$ is a numerical factor related to the degeneracy. Dividing
by a unit of volume in momentum space, $2\pi\hbar,$ gives the number
of states in phase space with the given energy, per unit volume of
phase space. Since we do not limit ourselves to nonrelativistic systems,
nor to three space dimensions, a more general definition is necessary
\cite{Kolekar,Paddy phase space}. The number of states is then defined
as\begin{equation}
N=\int d^{d}x\frac{d^{d}p}{(2\pi\hbar)^{d}}dE\delta(E-E(p)).\label{eq:N general first formula}\end{equation}
where $d$ denotes the number of space dimensions%
\footnote{This integral is actually $\int d^{d}x\frac{d^{d}p}{(2\pi\hbar)^{d+1}}dE\left[2\pi\hbar\delta(E-E(p))\right]$%. N as thus defined is divergent, and one must impose box normalization or a momentum cutoff.
}. Without loss of generality we are taking a constant time hypersurface.
The number of states is Lorentz invariant. For a proof see \cite{Paddy book}. 

Remarks on notation: for simplicity of notation in this paper, $g_{00}$
refers to the positive value of the time coordinate of the metric,
except where explicitly stated otherwise. The minus sign appears in
the form of the equation. An explicitly covariant derivation which
parallels ours can be found in \cite{Paddy phase space}. We here
keep the vector notation because it clarifies our proof in what follows.

In order to apply this definition to curved space, we need to clarify what momentum and energy refer to for a matter
field or gas of particles in curved space. There are (at least) two
possible ways to approach the issue. One is that of \cite{'t Hooft},
who took $\psi(x)$ a scalar wave function for a light spinless particle
of mass $m$ in the Schwarzschild metric, $m\ll1\ll M$ where $M$
is the BH mass, used a WKB approximation, wrote the wave equation
and defined the spatial momentum $k(r)$ in terms of the eigenvalues
of the Laplacian operator while taking energy as the eigenvalue of
the time component of the Laplacian. He obtained the number of states
by calculating $\int k(r)dr$ and then summing over angular degrees
of freedom. Another possibility is that of \cite{Kolekar,Paddy phase space}
who treated a relativistic gas of particles, and rather than the wave
equation, used the scalar invariance of the squared momentum four-vector
of the particle, while the covariant energy of a particle is the projection
of the timelike Killing vector on the four momentum. Both approaches
give the same relationship between energy and momentum, which for
a general static metric is \begin{equation}
g^{00}E^{2}=\sum_{i}g^{ii}\left(p_{i}\right)^{2}\label{eq:energy momentum eq}\end{equation}
taking a massless particle for simplicity. 

The number of states given by the volume of phase space
is the product of the volume of position and momentum space. The momentum
component of the number of states belongs to a constrained region
in the cotangent space of the region of configuration space in question.
For example, in Cartesian coordinates in flat space eq.(\ref{eq:energy momentum eq})
gives\begin{equation}
E^{2}-p_{x}^{2}-p_{y}^{2}-p_{z}^{2}=0\end{equation}
and this defines a sphere of radius $E$:\begin{equation}
1=\frac{p_{x}^{2}}{E^{2}}+\frac{p_{y}^{2}}{E^{2}}+\frac{p_{z}^{2}}{E^{2}}.\end{equation}
In statistical physics we take all energies up to a given energy,
and so we look for the volume enclosed by this sphere, $\frac{4}{3}\pi E^{3}$.
If the metric is not flat, the volume will be an ellipsoid. Since
our proof of curvature independence rests on a counter example, we
are free to take a static metric with timelike Killing vector and
can define the energy accordingly. 

For a general static diagonal metric eq. (\ref{eq:energy momentum eq})
gives \begin{eqnarray}
g^{00}E^{2} & = & \sum_{i}g^{ii}\left(p_{i}\right)^{2}\nonumber \\
1 & = & \sum_{i}\frac{g^{ii}\left(p_{i}\right)^{2}}{g^{00}E^{2}}=\sum_{i}\frac{p_{i}^{2}}{g_{ii}g^{00}E^{2}}\end{eqnarray}
where $p_{i}$ the spatial momenta are summed in all space directions.
This is the formula for an ellipsoid with axes $\sqrt{g_{ii}g^{00}}E,$
which encloses a region whose volume in three space dimensions would
be $\frac{4}{3}\pi\sqrt{g_{xx}g_{yy}g_{zz}}\left(\sqrt{g^{00}}E\right)^{3}$
. In $d+1$ spacetime dimensions this becomes \begin{equation}
C_{d}\sqrt{g_{d}}\left(\sqrt{g^{00}}E\right)^{d}\label{eq:ellipsoid}\end{equation}
where $g_{d}$ denotes the determinant of the spatial part of the
metric and $C_{d}$ is the volume enclosed by the d-dimensional unit
ball. One then integrates over all momentum space. Since the measure
in the momentum integral includes the root of inverse metric $g^{d}$,
that is, the integral is given by\begin{equation}
\int\frac{d^{d}p}{\sqrt{g_{d}}}\end{equation}
then the space determinant in eq.(\ref{eq:ellipsoid}) cancels out,
and the integral over momentum space gives the volume of a ball with
radius $\sqrt{g^{00}}E.$ 

Therefore the number of states in $d+1$ dimensions (d space dimensions)
for a diagonal metric is\begin{eqnarray}
N & = & C_{d}E^{d}\intop_{V}d^{d}x\sqrt{g_{d}}\left(g^{00}\right)^{\frac{d}{2}}\nonumber \\
C_{d} & = & \frac{\pi^{\frac{d}{2}}}{\Gamma\left(\frac{d}{2}+1\right)}.\end{eqnarray}
An explicit proof for $3+1$ and $4+1$ dimensions appears in the
Appendix.

\section{Invariance of number of states under transformation of metrics}

We wish to examine a general transformation which changes the metric
while leaving the number of states invariant. We find that such a
transformation exists, but does not preserve curvature. We give details
of the transformation, followed by examples of the relation to curvature.

We begin with conformal rescaling. If a $d-$dimensional metric changes
by $\tilde{g}_{\mu\nu}=a(x)g_{\mu\nu},$ then the number of states
is\begin{eqnarray}
N_{0} & = & \int\sqrt{g_{3}}d^{3}xd^{3}p=\int\sqrt{g_{3}}\frac{4\pi E^{3}}{3}\left(g^{00}\right)^{3/2}d^{3}x.\nonumber \\
\tilde{N} & = & \int\sqrt{\tilde{g}_{3}}d^{3}xd^{3}p\nonumber \\
 & = & \int a^{3/2}\sqrt{g}\frac{4\pi E^{3}}{3}\left(\frac{1}{a}g^{00}\right)^{3/2}d^{3}x\nonumber \\
 & = & \int\sqrt{g}\frac{4\pi E^{3}}{3}\left(g^{00}\right)^{3/2}d^{3}x=N.\end{eqnarray}
since $\tilde{g}_{00}=a(x)g_{00}$ and so $\tilde{g}^{00}=\frac{1}{a(x)}g^{00}$.This
only works if the metric is uniformly rescaled, so that $a_{0}=a_{i}.$
Thus conformal rescaling preserves the number of states. We conclude
that preservation of the number of states requires a constraint on
the relationship between the time and space components of the metric.

In search of a general transformation we take a general diagonal metric
in $1+3$ dimensions. Generalization to more space dimensions will
be simple.

\begin{equation}
\left(\begin{array}{cccc}
g_{00}\\
 & g_{xx}\\
 &  & g_{yy}\\
 &  &  & g_{zz}\end{array}\right)\end{equation}
The volume of space in this metric:\begin{equation}
\intop_{V}\sqrt{g_{xx}g_{yy}g_{zz}}d^{3}x\end{equation}
where the integral is over a given volume V. The volume of momentum
space is

\begin{equation}
\intop_{V_{p}}\frac{d^{3}p}{\sqrt{g_{xx}g_{yy}g_{zz}}}\end{equation}
where $V_{p}$ is the volume in momentum space. As explained above, from eq.(\ref{eq:energy momentum eq})
\begin{equation}
1=\frac{1}{g_{xx}g^{00}E^{2}}p_{x}^{2}+\frac{1}{g_{yy}g^{00}E^{2}}p_{y}^{2}+\frac{1}{g_{zz}g^{00}E^{2}}p_{z}^{2}\label{eq:wave eq}\end{equation}
which is the equation for volume of ellipsoid with axes $\sqrt{g_{xx}g^{00}}E,\sqrt{g_{yy}g^{00}}E,\sqrt{g_{zz}g^{00}E}$.
The momentum volume is obtained by integration, or more simply by
just plugging in the formula for volume of ellipsoid in 3 dimensions:
$\frac{4}{3}\pi abc=\frac{4}{3}\pi\sqrt{g_{xx}g_{yy}g_{zz}}\left(g^{00}E^{2}\right)^{3/2}.$

Phase space is given as%
\footnote{This can have a prefactor of $\left(2\pi\right)^{-3}$ when calculating
the density of modes per unit of phase space %
}:\begin{equation}
N=\intop_{V}d^{3}x\intop_{V_{p}}d^{3}p.\end{equation}
We now transform the metric in arbitrary way but keeping it diagonal:\begin{equation}
\left(\begin{array}{cccc}
a(\vec{x})g_{00}\\
 & b(\vec{x})g_{xx}\\
 &  & c(\vec{x})g_{yy}\\
 &  &  & d(\vec{x})g_{zz}\end{array}\right)\end{equation}
We plug this into the term for phase space. First we calculate the
volume of momentum space for the transformed metric. 
We obtain\begin{eqnarray}
\frac{1}{a(\vec{x})}g^{00}E^{2} & = & \frac{1}{b(\vec{x})}g^{xx}p_{x}^{2}+\frac{1}{c(\vec{x})}g^{yy}p_{y}^{2}+\frac{1}{d(\vec{x})}g^{zz}p_{z}^{2}\end{eqnarray}
 and using eq.(\ref{eq:wave eq})\begin{equation}
1=\frac{1}{a(\vec{x})b(\vec{x})g_{xx}g^{00}E^{2}}p_{x}^{2}+\frac{1}{a(\vec{x})c(\vec{x})g_{yy}g^{00}E^{2}}p_{y}^{2}+\frac{1}{a(\vec{x})d(\vec{x})g_{zz}g^{00}E^{2}}p_{z}^{2}\label{transf wave eq}\end{equation}
so that the volume becomes\begin{equation}
V_{p}=\frac{4}{3}\pi\sqrt{b(\vec{x})c(\vec{x})d(\vec{x})g_{xx}g_{yy}g_{zz}}\left(\frac{g^{00}}{a(\vec{x})}E^{2}\right)^{3/2}.\end{equation}
This will equal the volume before the transformation if\begin{equation}
b(\vec{x})c(\vec{x})d(\vec{x})=a(\vec{x})^{3}.\label{eq:THE CONSTRAINT}\end{equation}
Thus we have identified the constraint for an arbitrary transformation
to preserve the volume of phase space.

We looked for some kind of general algebraic characterization for
this kind of matrix, but found none. It belongs to $GL(n,R)$ but
does not represent a particular symmetry. Conformal transformations
are a subgroup of our transformation. Certain non conformal transformations
also preserve the number of states. This holds if the determinants
cancel out: That is, for $d$ space dimensions, the time part $a(x)$
when raised to the $d^{th}$ power, has to equal the determinant of
the space part. Take\begin{equation}
A=\left(\begin{array}{cccc}
a(x) & 0 & 0 & 0\\
0 & a(x)^{2} & 0 & 0\\
0 & 0 & a(x) & 0\\
0 & 0 & 0 & 1\end{array}\right)\end{equation}
As with the conformal transformation, we still have $\sqrt{\tilde{g}_{3}}=a^{3/2}\sqrt{g_{3}}$
and and so $\tilde{N}=N_{0}$. 

So in general our constraint is:\begin{equation}
\left(g_{00}\right)^{d}=det\: g_{space}\label{eq:constraint in GENERAL}\end{equation}
where $d$ is the number of space dimensions and $g_{space}$ is the
determinant of the spatial part of the metric.

We can regard the transformation matrix, labeled $A$, as two blocks,
separating the time and space components:\begin{equation}
A=\left(\begin{array}{cc}
T\\
 & S\end{array}\right)\end{equation}
where $T$ is a 1x1 matrix, and S is a diagonal matric of rank $d$
where $d$ is the dimension of space (rank of $A$ is the dimension
of space-time, $d+1$). Then the constraint requires \begin{eqnarray}
det(S) & = & det(T)^{d}\nonumber \\
det(A) & = & det(T)^{2d}.\end{eqnarray}

\section{Relation to curvature}

In $d+1$ dimensions \begin{eqnarray}
N & \sim & \int d^{3}x\sqrt{\frac{g_{d}}{\left(g_{00}\right)^{d}}}\end{eqnarray}
where $g_{d}$ denotes the determinant of the space part of metric.
To preserve the number of states we have to preserve the ratio $g_{d}/g_{00}^{d}$
, which entails the constraint on the determinant as detailed above.
The question becomes: given a change of metric for which this constraint
holds, will such a constraint ensure preservation of scalar curvature?
If so preservation of the number of states would entail preservation
of curvature, which is an observer independent characteristic.

We take two matrices representing two possible transformations of
a given $3$- dimensional metric:\begin{eqnarray}
A & = & \frac{1}{L}\left(\begin{array}{ccc}
x\\
 & x\\
 &  & x\end{array}\right)\nonumber \\
B & = & \left(\begin{array}{ccc}
\frac{\sqrt{2}x}{L}\\
 & 2\\
 &  & \frac{x^{2}}{L^{2}}\end{array}\right)\end{eqnarray}
where $L$ is a constant with dimension of length. Both transformation
matrices preserve the constraint given above, while their curvature
differs. That is, taking a flat metric for example, after undergoing each of these transformations it would have the same number of states as previously, but different curvature. The first transformation would give  $R=\frac{3L}{2x^{3}}$ , the secondgives $R=\frac{L}{\sqrt{2}x^{3}}.$
This is because the second one has fewer Christoffel signs, since
the derivative must be $\partial_{x}$, and $\partial_{x}g_{yy}=0$
. Therefore clearly imposing the constraint on a metric transformation
will not necessarily preserve the curvature of the original metric.

Curvature in these examples is affected by the number of terms with
an $x$ derivative, while the determinant is not. Thus the constraint
on the determinant does NOT preserve curvature. This is intuitively
understandable: the determinant indicates volume but gives no information
as to the spatial distribution of the volume.

\subsection{Examples in various dimensions}

In $1+1$ dimensions a transformation that preserves $N$ \emph{must}
be conformal: $g_{00}=g_{xx}$ since $det\, g_{space}=g_{xx.}$ In
2+1 dimensions we give two examples of transformations that preserve
N. One is conformal, the other non conformal but symmetric: 

Conformal: \begin{equation}
A=\frac{1}{L}\left(\begin{array}{ccc}
x\\
 & x\\
 &  & x\end{array}\right),R=\frac{3L}{2x^{3}}\end{equation}

Symmetric:\begin{equation}
B=\frac{1}{L}\left(\begin{array}{ccc}
\sqrt{xy}\\
 & x\\
 &  & y\end{array}\right),R=L\left(\frac{5x^{2}-6y\sqrt{xy}}{8x^{3}y^{2}}\right).\end{equation}
An asymmetric example is like the one given in the previous section
for a Euclidean metric.

Note that plugging in the value $x=y$ \emph{after} deriving $R$
for matrix $B$ does not give the curvature of matrix $A.$ This is
because the derivation of $R$ takes into account the direction of
each component as well as its numerical value. If one plugs in $y=x$
before deriving $R$ all the derivatives $\partial_{y}$ vanish, giving
the different result.

We next look at 3+1 dimensions. The constraint requires $\left|\left(g_{00}\right)^{3}\right|=detg_{3}$.
Comparing several matrices that obey this constraint and inspecting
their curvature:\begin{eqnarray}
A=\left(\begin{array}{cccc}
\frac{x}{L}\\
 & \frac{x^{3}}{L^{3}}\\
 &  & 1\\
 &  &  & 1\end{array}\right),\: R & = & \frac{2L^{3}}{x^{5}}\nonumber \\
B=\frac{1}{L}\left(\begin{array}{cccc}
\left(xyz\right)^{\frac{1}{3}}\\
 & x\\
 &  & y\\
 &  &  & z\end{array}\right),\: R & = & \frac{4L}{9}\left(\frac{1}{x^{3}}+\frac{1}{y^{3}}+\frac{1}{z^{3}}\right)\nonumber \\
C=\frac{1}{L}\left(\begin{array}{cccc}
x\\
 & x\\
 &  & x\\
 &  &  & x\end{array}\right),\: R & = & \frac{3L}{2x^{3}}\end{eqnarray}
A few comments: 1) The curvature for the third transformation is the
same as for the conformal matrix in 1+2 dimensions. 2) As before,
setting $x=y=z$ after calculating the curvature for matrix $B$ does
not give the same result as the curvature for matrix $C$. Again,
this is because the direction of the variable contributes in calculating
$R,$ and not just its numeric value. This sheds light on the fact
that the number of states, which is proportional to the volume of
phase space, is different from curvature, which incorporates information
on the distribution of that volume.\textsf{\textbf{\textit{ }}}A constraint
on the determinant, representing Euclidean volume, is not the same
as that on Ricci curvature, which in fact represents the amount by
which the volume of a geodesic ball in a curved Riemannian manifold
deviates from that of the standard ball in Euclidean space.

\subsubsection{Rindler vs Schwarzschild: }

The transformation from Minkowsky to Rindler space is not diagonal.
It mixes time and space coordinates and that is why N is different
from flat space. We cannot conclude from this that curvature is irrelevant
to statistical entropy. That conclusion can only be drawn from the
general proof given above.

The Schwarzschild metric diverges at the boundary and it was found
that the number of states (and thus the entropy) is different from
that of Minkowski space \cite{'t Hooft,JudyRamy}. This is not the
same as the difference between the number of states in Rindler and
Minkowski spaces. In the Schwarzschild case the argument above does
apply, since the transformation metric from Minkowski to Schwarzschild
metric is diagonal. The Schwarzschild number of states differs from
that of Minkowski because of the redshift on energy: $g_{00}(r)$.

\subsection{Discussion}

Our transformation leaves N invariant because it preserves the relationship
between the volume of momentum space and of position space. $\left(g^{00}\right)^{3/2}$
is the variable part of momentum space, and $\sqrt{g_{d}}$ is the
variable part of position space. $N$ is invariant so long as the
relation between the two is preserved, so that if position space shrinks,
momentum space grows and vice verse: $a(x)^{d}$ multiplying $\sqrt{g_{space}}$
equals $1/a(x)^{d}$ multiplying momentum space.

We examined the question whether in curved space the number of states,
and the statistical entropy derived from this, is observer dependent
or is related to a physical quantity such as curvature. We found that
it is observer dependent and not related to the intrinsic geometry. 

One might argue that a proof of curvature independence must show that
there are no cases at all where curvature is preserved under a transformation
that preserved the number of states. In fact it is quite possible
that in some case curvature might be preserved. We claim that this
must be seen as a coincidence because the constraint on preservation
of the number of states relates to the determinant. By definition,
there is a difference between the determinant, which represents Euclidean
volume and does not depend on directions in space, and curvature which
does depend on directions in space. The number of states does not
depend on directions in space and so it can be preserved even if the
directional characteristics and thus the curvature are changed. There
will be a subgroup where transformation of the number of states will
indeed preserve curvature. But one cannot assume that any given number
of states, and the entropy derived from it, relate to a spacetime
with a given curvature.

The results in this paper apply to a diagonal metric only, but this
is sufficient as it serves as a counter example. The question arises:
what of Wald's entropy? Since statistical entropy is derived from
the number of states, whereas Wald's entropy is an explicit function
of curvature, this indicates a difference between these two concepts
of entropy. Calculation for Einstein gravity gives the same result
in both cases, but for generalized theories of gravity Wald entropy
could contain terms derived from the curvature, so the concepts themselves
do not coincide.

However the issue may not be so simple. Phase space is defined as
the product of the volume of position and momentum spaces. The definition
arose in the context where momentum refers to kinetic momentum which
is also the canonical conjugate to position. However even in spherical
coordinates, kinetic and conjugate momentum do not coincide \cite{Chinese},
and a treatment in curved space should generalize the definition to
conjugate momentum. If this is done, one then notes that gravitational
Lagrangian includes the Ricci scalar, and Ricci tensors as well in
the generalized theories of gravity with which Wald dealt. The Lagrangian
of a particle in a gravitational background will include at least
two terms, the matter Lagrangian and the graviational term. Each will
have a generalized momentum conjugate to the dynamical variable in
the Lagrangian. Therefore it may be necessary to redefine statistical
entropy to take into account a more general formulation of phase space.
This cannot be done simply by adding gravitational degrees of freedom;
for example, adding gravitational degrees of freedom to the statistical
calculation would not give one fourth the area but one half, and thus
differ from the other derivations of entropy. A clue to suitable redefinition of the term may be found in the example discussed at the start of the paper: Minkowski and Rindler space. Statistical entropy was originally defined for flat space, where choice of vacuum is unambiguous. In treatment of curved space there should be a way to incorporate the choice of vacuum into the concept of phase space.

Another issue is that of divergence on the horizon. While this may
be an artifact of quantum uncertainty\cite{JudyRamy} still a more
thorough investigation is necessary before drawing conclusions on
the relationship of the two entropies. 

Note: There is a claim that entanglement entropy and statistical entropy
are one and the same. In \cite{Katja} it was shown for explicit examples
that entanglement entropy does not depend on curvature. For a discretized
region in curved space it was found that even when the space is large
enough for the effects of curvature to be noticeable, entropy remains
proportional to area and is not affected by the curvature of the background.
This qualitative similarity to our result reinforces the idea that
entanglement and statistical entropy may be the same, and that they
differ from Noether charge entropy.

In conclusion, we have shown that the number of states is a function
of the metric and is preserved under specific transformations of the
metric, which do not necessarily preserve curvature. Therefore the
number of states calculated with the accepted definition of phase
space does not depend on curvature, and neither does the statistical
entropy derived from it. For general theories of gravity it may be
necessary to redefine statistical entropy taking into account a more
general concept of phase space in some subtle manner, but as the definitions
stand, it appears that statistical entropy and Wald entropy differ

This research was supported by the Israel Science Foundation Grant
No. 239/10. We thank Ramy Brustein, Merav Hadad and Frol Zapolsky
for helpful discussions, and Joey Medved for comments on the manuscript.

\appendix

\section{Phase space in 3+\label{sec:Phase space volume}1 and in 4+1 dimensions}

The number of states in $d+1$ dimensions (d space dimensions) for
a diagonal metric works out to be\begin{eqnarray}
N & = & CE^{d}\intop_{V}d^{d}x\sqrt{g_{d}}\left(g^{00}\right)^{\frac{d}{2}}\nonumber \\
C & = & \frac{\pi^{\frac{d}{2}}}{\Gamma\left(\frac{d}{2}+1\right)}\end{eqnarray}
and $g_{d}$ is the determinant of the spatial components of the metric.

\subsection*{Proof in 3+1 dimensi\label{sub:Proof-in-3+1dimensions}ons:}

 Eq.(\ref{eq:energy momentum eq}) gives:\begin{eqnarray*}
g^{00}E^{2}-g^{xx}p_{x}^{2}-g^{yy}p_{y}^{2}-g^{zz}p_{z}^{2} & = & 0\end{eqnarray*}
 \[
p_{x}=\sqrt{g_{xx}}\sqrt{g^{00}E^{2}-g^{yy}p_{y}^{2}-g^{zz}p_{z}^{2}}\]
\begin{eqnarray*}
\intop d^{3}p & = & \intop_{-\sqrt{g_{yy}g^{00}}E}^{\sqrt{g_{yy}g^{00}}E}dp_{y}\intop_{-\sqrt{g_{zz}}\sqrt{g^{00}E^{2}-g^{yy}p_{y}^{2}}}^{\sqrt{g_{zz}}\sqrt{g^{00}E^{2}-g^{yy}p_{y}^{2}}}dp_{z}\intop_{-\sqrt{g_{xx}}\sqrt{g^{00}E^{2}-g^{yy}p_{y}^{2}-g^{zz}p_{z}^{2}}}^{\sqrt{g_{xx}}\sqrt{g^{00}E^{2}-g^{yy}p_{y}^{2}-g^{zz}p_{z}^{2}}}dp_{x}\\
 & = & 2\sqrt{g_{xx}}\intop_{-\sqrt{g_{yy}g^{00}}E}^{\sqrt{g_{yy}g^{00}}E}dp_{y}\intop_{-\sqrt{g_{zz}}\sqrt{g^{00}E^{2}-g^{yy}p_{y}^{2}}}^{\sqrt{g_{zz}}\sqrt{g^{00}E^{2}-g^{yy}p_{y}^{2}}}dp_{z}\,\sqrt{g^{00}E^{2}-g^{yy}p_{y}^{2}-g^{zz}p_{z}^{2}}\end{eqnarray*}
We label $g^{00}E^{2}-g^{yy}p_{y}^{2}\equiv A^{2}$. Then the integral
over $p_{z}$ becomes\begin{eqnarray}
\intop_{-\sqrt{g_{zz}}A}^{\sqrt{g_{zz}}A}dp_{z}\sqrt{A^{2}-g^{zz}p_{z}^{2}} & = & A\intop_{-\sqrt{g_{zz}}A}^{\sqrt{g_{zz}}A}dp_{z}\sqrt{1-\frac{p_{z}^{2}}{g_{zz}A^{2}}}\nonumber \\
 & = & A^{2}\sqrt{g_{zz}}\intop_{-1}^{1}du\sqrt{1-u^{2}}=A^{2}\sqrt{g_{zz}}\frac{\pi}{2}.\end{eqnarray}
Plugging this in we get\begin{eqnarray}
\int d^{3}p & = & 2\sqrt{g_{xx}g_{zz}}\frac{\pi}{2}\intop_{-\sqrt{g_{yy}g^{00}}E}^{\sqrt{g_{yy}g^{00}}E}dp_{y}\left(g^{00}E^{2}-g^{yy}p_{y}^{2}\right)\nonumber \\
 & = & \sqrt{g_{xx}g_{zz}}\pi\left[2\sqrt{g_{yy}}\left(g^{00}E^{2}\right)^{3/2}-\frac{2}{3}g^{yy}\left(\sqrt{g_{yy}g^{00}}E\right)^{3}\right]\nonumber \\
 & = & \sqrt{g_{xx}g_{zz}g_{yy}}\frac{4}{3}\pi\left(\sqrt{g^{00}}E\right)^{3}.\end{eqnarray}

\subsection*{One more dimension:}

\begin{eqnarray*}
g^{00}E^{2}-g^{xx}p_{x}^{2}-g^{yy}p_{y}^{2}-g^{zz}p_{z}^{2}-g^{ww}p_{w}^{2} & = & 0\\
p_{w}=\sqrt{g_{ww}}\sqrt{g^{00}E^{2}-g^{xx}p_{x}^{2}-g^{yy}p_{y}^{2}-g^{zz}p_{z}^{2}}\end{eqnarray*}

\textsf{\textbf{\textit{\begin{eqnarray}
\intop d^{3}p & = & \intop_{-\sqrt{g_{yy}g^{00}}E}^{\sqrt{g_{yy}g^{00}}E}dp_{y}\intop_{-\sqrt{g_{zz}}\sqrt{g^{00}E^{2}-g^{yy}p_{y}^{2}}}^{\sqrt{g_{zz}}\sqrt{g^{00}E^{2}-g^{yy}p_{y}^{2}}}dp_{z}\intop_{-\sqrt{g_{xx}}\sqrt{g^{00}E^{2}-g^{yy}p_{y}^{2}-g^{zz}p_{z}^{2}}}^{\sqrt{g_{xx}}\sqrt{g^{00}E^{2}-g^{yy}p_{y}^{2}-g^{zz}p_{z}^{2}}}dp_{x}\times\nonumber \\
 &  & \times\intop_{-\sqrt{g_{ww}}\sqrt{g^{00}E^{2}-g^{xx}p_{x}^{2}-g^{yy}p_{y}^{2}-g^{zz}p_{z}^{2}}}^{\sqrt{g_{ww}}\sqrt{g^{00}E^{2}-g^{xx}p_{x}^{2}-g^{yy}p_{y}^{2}-g^{zz}p_{z}^{2}}}dp_{w}\\
 & = & 2\sqrt{g_{ww}}\intop_{-\sqrt{g_{yy}g^{00}}E}^{\sqrt{g_{yy}g^{00}}E}dp_{y}\intop_{-\sqrt{g_{zz}}\sqrt{g^{00}E^{2}-g^{yy}p_{y}^{2}}}^{\sqrt{g_{zz}}\sqrt{g^{00}E^{2}-g^{yy}p_{y}^{2}}}dp_{z}\times\nonumber \\
 &  & \times\intop_{-\sqrt{g_{xx}}\sqrt{g^{00}E^{2}-g^{yy}p_{y}^{2}-g^{zz}p_{z}^{2}}}^{\sqrt{g_{xx}}\sqrt{g^{00}E^{2}-g^{yy}p_{y}^{2}-g^{zz}p_{z}^{2}}}dp_{x}\sqrt{g^{00}E^{2}-g^{xx}p_{x}^{2}-g^{yy}p_{y}^{2}-g^{zz}p_{z}^{2}}\end{eqnarray}
}}}We label $g^{00}E^{2}-g^{zz}p_{z}^{2}-g^{yy}p_{y}^{2}\equiv A^{2}$.
Then the integral over $p_{x}$ becomes\begin{eqnarray}
\intop_{-\sqrt{g_{xx}}A}^{\sqrt{g_{xx}}A}dp_{x}\sqrt{A^{2}-g^{xx}p_{x}^{2}} & = & A\intop_{-\sqrt{g_{xx}}A}^{\sqrt{g_{xx}}A}dp_{x}\sqrt{1-\frac{p_{x}^{2}}{g_{xx}A^{2}}}\nonumber \\
 & = & A^{2}\sqrt{g_{xx}}\intop_{-1}^{1}du\sqrt{1-u^{2}}=A^{2}\sqrt{g_{xx}}\frac{\pi}{2}.\end{eqnarray}
Plugging this in we get\begin{eqnarray*}
\int d^{3}p & = & 2\sqrt{g_{xx}g_{ww}}\frac{\pi}{2}\intop_{-\sqrt{g_{yy}g^{00}}E}^{\sqrt{g_{yy}g^{00}}E}dp_{y}\intop_{-\sqrt{g_{zz}}\sqrt{g^{00}E^{2}-g^{yy}p_{y}^{2}}}^{\sqrt{g_{zz}}\sqrt{g^{00}E^{2}-g^{yy}p_{y}^{2}}}dp_{z}\left(g^{00}E^{2}-g^{yy}p_{y}^{2}-g^{zz}p_{z}^{2}\right)\end{eqnarray*}
Let us label $g^{00}E^{2}-g^{yy}p_{y}^{2}\equiv B^{2}.$ Then the
$p_{z}$integral becomes\begin{equation}
\intop_{-\sqrt{g_{zz}}B}^{\sqrt{g_{zz}}B}dp_{z}\left(B^{2}-g^{zz}p_{z}^{2}\right)=\frac{4}{3}\sqrt{g_{zz}}B^{3}=\frac{4}{3}\sqrt{g_{zz}}\left(g^{00}E^{2}-g^{yy}p_{y}^{2}\right)^{3/2}.\end{equation}
Integrate over $p_{y}$:\begin{eqnarray}
\intop_{-\sqrt{g_{yy}g^{00}}E}^{\sqrt{g_{yy}g^{00}}E}dp_{y}\left(g^{00}E^{2}-g^{yy}p_{y}^{2}\right)^{3/2} & = & \sqrt{g_{yy}}\frac{3}{8}\pi\left(\sqrt{g^{00}}E\right)^{4}.\end{eqnarray}
(this was done with Mathematica, you get a result containing $Arctan[\infty]=\frac{\pi}{2}$).

Plugging this back in,\begin{eqnarray*}
\int d^{3}p & = & \frac{4}{3}\left(\frac{3}{8}\right)\pi^{2}\sqrt{g_{xx}g_{ww}g_{yy}g_{zz}}\left(g^{00}\right)^{2}E^{4}\\
 & = & \frac{\pi^{2}}{2}\sqrt{g_{4}}\left(g^{00}\right)^{2}E^{4}\end{eqnarray*}
and so\begin{eqnarray}
N & =\frac{\pi^{2}}{2}E^{4} & \intop_{V}d^{4}x\sqrt{g_{4}}\left(g^{00}\right)^{2}\nonumber \\
 & = & C_{d}E^{4}\intop_{V}d^{4}x\sqrt{g_{4}}\left(g^{00}\right)^{\frac{4}{2}},\:\left(C=\frac{\pi^{\frac{4}{2}}}{\Gamma\left(\frac{4}{2}+1\right)}\right)\end{eqnarray}
just as we claimed. We would like to be able to prove by induction
that if it's true for $N_{d}$ it's true for $N_{d+1}$ but (so far)
we can't generalize the integration: the integrand becomes $\left(g^{00}E^{2}-g^{(d+1,d+1)}p_{d+1}\right)^{d/2}.$


\begin{thebibliography}{24}
\bibitem{BCH}J.M.Bardeen, B.Carter and S.W.Hawking, Commun.Math.Phys.31,161
(1973).

\bibitem{Bekenstein}J.D.Bekenstein, Phys. Rev. D 7, 2333 (1973).

\bibitem{Gibbons hawking}G.W. Gibbons and S.W. Hawking, , Phys. Rev.
D, 15, 2752\textendash{}2756, (1977).

\bibitem{'t Hooft}G.\textquoteright{}t Hooft, Nucl. Phys. B256, 727
(1985). 

\bibitem{Paddy}T. Padmanabhan, Rep. Prog. Phys. 73,046901 (2010)
{[}arXiv:0911.5004v2{]} and references therein. 

\bibitem{Wald review}R. M. Wald, Living Rev. Relativ. 4, 6 (2001).
http://www.livingreviews. org/lrr-2001-6, {[}arXiv:gr-qc/9912119v2{]}.

\bibitem{Bombelli}Luca Bombelli, Rabinder K. Koul, Joohan Lee, and
Rafael D. Sorkin, , Phys. Rev. D 34, 373 (1986).

\bibitem{Srednicki}Mark Srednicki, Phys. Rev. Lett. 71, 666 (1993).

\bibitem{Plenio}M.B.Plenio, J. Eisert, J. Dreissig and M. Cramer,
Phys. Rev. Lett. 94, 060503 (2005) {[}arXiv:quant-ph/0405142v3 {]}.
See also J. Eisert, M. Cramer, and M. B. Plenio, Rev. Mod. Phys. 82,
277 (2010) {[}arXiv:0808.3773v4{]}.

\bibitem{Wald 1}Robert M.Wald, Phys. Rev. D 48, 3427-3431 (1993)
{[}arXiv:gr-qc/9307038v1{]}.

\bibitem{wald iyer}V. Iyer and R.M.Wald, Phys. 4 Rev. D 50, 846 (1994)
{[}arXiv:gr-qc/9403028v1{]}.

\bibitem{Carlip}S.Carlip, Class.Quant.Grav.16:3327-3348 (1999) {[}arXiv:gr-qc/99061262v2{]}.

\bibitem{Silva}S. Silva, Class.Quant.Grav. 19 (2002) 3947-3962 {[}hep-th/0204179{]}.

\bibitem{Unruh}W.G. Unruh, Phys. Rev. D 14 (4), 870 (1976).

\bibitem{paddy kolekar ref 13}T.Padmanabhan, 2010 Mod. Phys. Lett.
A 25 1129 (2010) {[}arXiv:0912.3165{]}; Phys. Rev. D 81 124040 (2010)
{[}arXiv:1003.5665{]};

\bibitem{Kolekar}Sanved Kolekar, T.Padmanabhan, Phys. Rev.D83, 064034
(2011) {[}arXiv:1012.5421{]}

\bibitem{Kabat}D.Kabat, Nuclear Physics B 453, 281 (1995) {[}arXiv:hep-th/9503016{]}

\bibitem{kabat 3 israel}W. Israel, Phys. Lett. A57, 107 (1976).

\bibitem{JudyRamy}R.Brustein and J.Kupferman, Phys. Rev.D83, 124014
(2011) {[}arXiv:1010.4157v2{]}.

\bibitem{Paddy phase space}T. Padmanabhan. Physics Letters A, 136,203\textendash{}205,
(1989).

\bibitem{Paddy book}T. Padmanabhan, \emph{Gravitation: Foundations
and Frontiers}, Cambridge University Press, (2010), p.36.

\bibitem{Chinese}Lee Ting Hsang, An Chong Shan and Zhai Tian Yi,
International Journal of Theoretical Physics 29, 9 (1990)

\bibitem{Katja}Katja Ried, quant-ph/1309.7380v1 (2013).

\end{thebibliography}
\end{document}